\documentstyle[twoside,fleqn]{article}
\input{epsf}

\catcode`\@=11

\def\fileversion{v2.6}
\def\filedate{24 November 1993}

\newdimen\@bls                    
\@bls=\baselineskip               
\advance\@bls -1ex                
\newdimen\@eps                    %
\def\section{\@startsection{section}{1}{\z@}
  {1.5\@bls plus 0.5\@bls}{1\@bls}{\normalsize\bf}}
\def\subsection{\@startsection{subsection}{2}{\z@}
  {1\@bls plus 0.25\@bls}{\@eps}{\normalsize\bf}}
\def\subsubsection{\@startsection{subsubsection}{3}{\z@}
  {1\@bls plus 0.25\@bls}{\@eps}{\normalsize\bf}}
\def\paragraph{\@startsection{paragraph}{4}{\parindent}
  {1\@bls plus 0.25\@bls}{0.5em}{\normalsize\bf}}
\def\subparagraph{\@startsection{subparagraph}{4}{\parindent}
  {1\@bls plus 0.25\@bls}{0.5em}{\normalsize\bf}}

\def\@sect#1#2#3#4#5#6[#7]#8{\ifnum #2>\c@secnumdepth
  \def\@svsec{}\else 
  \refstepcounter{#1}\edef\@svsec{\csname the#1\endcsname.\hskip0.5em}\fi
  \@tempskipa #5\relax
  \ifdim \@tempskipa>\z@
    \begingroup 
      #6\relax
      \@hangfrom{\hskip #3\relax\@svsec}{\interlinepenalty \@M #8\par}%
    \endgroup
    \csname #1mark\endcsname{#7}\addcontentsline
      {toc}{#1}{\ifnum #2>\c@secnumdepth \else
        \protect\numberline{\csname the#1\endcsname}\fi #7}%
  \else
    \def\@svsechd{#6\hskip #3\@svsec #8\csname #1mark\endcsname
      {#7}\addcontentsline{toc}{#1}{\ifnum #2>\c@secnumdepth \else
        \protect\numberline{\csname the#1\endcsname}\fi #7}}%
  \fi \@xsect{#5}}

\long\def\@makefigurecaption#1#2{\vskip 10mm #1. #2\par}

\long\def\@maketablecaption#1#2{\hbox to \hsize{\parbox[t]{\hsize}
  {#1 \\ #2}}\vskip 0.3ex}

\def\fnum@figure{Figure \thefigure}
\def\figure{\let\@makecaption\@makefigurecaption \@float{figure}}
\@namedef{figure*}{\let\@makecaption\@makefigurecaption \@dblfloat{figure}}

\def\table{\let\@makecaption\@maketablecaption \@float{table}}
\@namedef{table*}{\let\@makecaption\@maketablecaption \@dblfloat{table}}

\textfloatsep=\floatsep           
\intextsep=\floatsep              

\long\def\@makefntext#1{\parindent 1em\noindent\hbox{${}^{\@thefnmark}$}#1}

\def\maketitle{\begingroup        
    \def\thefootnote{\fnsymbol{footnote}}%
    \newpage \global\@topnum\z@ 
    \@maketitle \@thanks
  \endgroup
  \let\maketitle\relax \let\@maketitle\relax
  \gdef\@thanks{}\let\thanks\relax
  \gdef\@address{}\gdef\@author{}\gdef\@title{}\let\address\relax}

\def\justify@on{\let\\=\@normalcr
  \leftskip\z@ \@rightskip\z@ \rightskip\@rightskip}

\newbox\fm@box                    

\def\@maketitle{
  \global\setbox\fm@box=\vbox\bgroup
    \vskip 8mm                    
    \raggedright                  
    \hyphenpenalty\@M             
    {\Large \@title \par}         
    \vskip\@bls                   
    {\normalsize                  
     \@author \par}               
    \vskip\@bls                   
    \@address                     
  \egroup
  \twocolumn[
    \unvbox\fm@box                
    \vskip\@bls                   
    \unvbox\abstract@box          
    \vskip 2pc]}                  

\newcounter{address} 
\def\theaddress{\alph{address}}
\def\@makeadmark#1{\hbox{$^{\rm #1}$}}   

\def\address#1{\addressmark\begingroup
  \xdef\@tempa{\theaddress}\let\\=\relax
  \def\protect{\noexpand\protect\noexpand}\xdef\@address{\@address
  \protect\addresstext{\@tempa}{#1}}\endgroup}
\def\@address{}

\def\addressmark{\stepcounter{address}%
  \xdef\@tempb{\theaddress}\@makeadmark{\@tempb}}

\def\addresstext#1#2{\leavevmode \begingroup
  \raggedright \hyphenpenalty\@M \@makeadmark{#1}#2\par \endgroup
  \vskip\@bls}

\newbox\abstract@box              

\def\abstract{%
  \global\setbox\abstract@box=\vbox\bgroup
  \small\rm
  \ignorespaces}
\def\endabstract{\par \egroup}

\def\thebibliography#1{\section*{REFERENCES}\list{\arabic{enumi}.}
  {\settowidth\labelwidth{#1.}\leftmargin=1.67em
   \labelsep\leftmargin \advance\labelsep-\labelwidth
   \itemsep\z@ \parsep\z@
   \usecounter{enumi}}\def\makelabel##1{\rlap{##1}\hss}%
   \def\newblock{\hskip 0.11em plus 0.33em minus -0.07em}
   \sloppy \clubpenalty=4000 \widowpenalty=4000 \sfcode`\.=1000\relax}

\newcount\@tempcntc
\def\@citex[#1]#2{\if@filesw\immediate\write\@auxout{\string\citation{#2}}\fi
  \@tempcnta\z@\@tempcntb\m@ne\def\@citea{}\@cite{\@for\@citeb:=#2\do
    {\@ifundefined
       {b@\@citeb}{\@citeo\@tempcntb\m@ne\@citea
        \def\@citea{,\penalty\@m\ }{\bf ?}\@warning
       {Citation `\@citeb' on page \thepage \space undefined}}%
    {\setbox\z@\hbox{\global\@tempcntc0\csname b@\@citeb\endcsname\relax}%
     \ifnum\@tempcntc=\z@ \@citeo\@tempcntb\m@ne
       \@citea\def\@citea{,\penalty\@m}
       \hbox{\csname b@\@citeb\endcsname}%
     \else
      \advance\@tempcntb\@ne
      \ifnum\@tempcntb=\@tempcntc
      \else\advance\@tempcntb\m@ne\@citeo
      \@tempcnta\@tempcntc\@tempcntb\@tempcntc\fi\fi}}\@citeo}{#1}}

\def\@citeo{\ifnum\@tempcnta>\@tempcntb\else\@citea
  \def\@citea{,\penalty\@m}%
  \ifnum\@tempcnta=\@tempcntb\the\@tempcnta\else
   {\advance\@tempcnta\@ne\ifnum\@tempcnta=\@tempcntb \else
\def\@citea{--}\fi
    \advance\@tempcnta\m@ne\the\@tempcnta\@citea\the\@tempcntb}\fi\fi}

\def\ps@crcplain{\let\@mkboth\@gobbletwo
	\def\@oddhead{\box10\hfil}
	\def\@evenhead{\hfil}
     \def\@oddfoot{\hfil\rm\thepage\hfil}
     \def\@evenfoot{\hfil\rm\thepage\hfil}}

\ps@crcplain                    

\catcode`\@=12

\newcommand{\slL}{\raise.15ex\hbox{$/$}\kern-.53em\hbox{$L$}}
\newcommand{\slP}{\raise.15ex\hbox{$/$}\kern-.53em\hbox{$P$}}
\newcommand{\slR}{\raise.15ex\hbox{$/$}\kern-.53em\hbox{$R$}}
\newcommand{\slQ}{\raise.15ex\hbox{$/$}\kern-.53em\hbox{$Q$}}

\catcode`\@=11
\def\eqalign#1{\null\vbox{\openup\jot\m@th
\halign to\linewidth{
\tabskip=0mm plus 30mm\strut$\displaystyle{##}$&\hfill${##}$\tabskip=0mm\crcr #1\crcr}}}

\def\cno{\global\advance\c@equation by 1\relax
	(\the\c@equation)}

\font\cmr=cmr7

\font\tenimbf=cmmib10 at 12pt
\font\sevenimbf=cmmib10 at 7pt
\font\fiveimbf=cmmib10 at 5pt
\newfam\imbf
\textfont\imbf=\tenimbf
\scriptfont\imbf=\sevenimbf
\scriptscriptfont\imbf=\fiveimbf

\font\tenmsa=msam10
\font\sevenmsa=msam7
\font\fivemsa=msam5
\font\tenmsb=msbm10
\font\sevenmsb=msbm7
\font\fivemsb=msbm5
\newfam\msafam
\newfam\msbfam
\textfont\msafam=\tenmsa  \scriptfont\msafam=\sevenmsa
  \scriptscriptfont\msafam=\fivemsa
\textfont\msbfam=\tenmsb  \scriptfont\msbfam=\sevenmsb
  \scriptscriptfont\msbfam=\fivemsb

\title{Enhanced photon production near the light--cone 
by a hot plasma
\thanks{Work done in collaboration with P.~Aurenche, R.~Kobes, 
and E.~Petitgirard.}
\thanks{Talk given at the
QCD96 conference, Montpellier, 4-12th July 1996.}}

\author{ F. Gelis\address{Laboratoire de Physique Th\'eorique ENSLAPP,\\
 URA 14-36 du CNRS, associ\'ee \`a l'Ecole Normale Sup\'erieure de Lyon
et \`a l'Universit\'e de Savoie,\\
B.P. 110, F-74941 Annecy-le-Vieux Cedex, France}
}

\begin{document}

\begin{abstract}
Strong collinear divergences, although regularized by a thermal mass,
result in a breakdown of the standard hard thermal loop expansion
in the calculation of the production rate of photons by a plasma
of quarks and gluons using thermal field theory techniques.
\end{abstract}

\maketitle

\section{INFRARED AND COLLINEAR PROBLEMS AT FINITE T}
\subsection{Hard thermal loop resummation}
In this talk, we study the production of soft
photons by a hot quark gluon plasma. Indeed, this
is a good way of testing our understanding
of the infrared and collinear behavior of thermal gauge theories.
Intuitively, these difficulties are increased with respect to
the $T=0$ case, because of the presence in the Feynman rules
of Bose-Einstein factors $n_{_{B}}(k_0)=1/(\exp(k_0/T)-1)$
which can be large if $k_0$ is much smaller than the 
temperature $T$. 
Concerning the collinear singularities, they can no longer be 
factorized in the hadronic structure functions since we are now in 
a deconfined phase.

To improve the infrared behavior of thermal field theories, a
resummation, known as the hard thermal loop (HTL) resummation, has
been proposed by \cite{htl}. Basically, this resummation 
lies in the fact that certain thermal radiative corrections 
can be as large as the bare corresponding 
quantity. In particular, this is the case when the external
legs of a propagator or vertex carry only soft momenta (i.e. of
order $gT$). 
\nopagebreak
\subsection{Effective propagators and vertices}
In QCD, we need effective gluon and quark propagators, as well
as effective vertices. The resulting effective theory turns out
to be gauge invariant. For later reference, let us quote the effective
gluon propagator:
\begin{equation}
D^{\mu\nu}(L)=\!\!\!\!\!\sum\limits_{\alpha=T,L}\!
{{-iP^{\mu\nu}_{\alpha}(L)}
\over{L^2-\Pi_{\alpha}(L)+i\epsilon}}+
{{i\xi P^{\mu\nu}_{_{G}}(L)}\over{L^2+i\epsilon}},
\end{equation}
in a covariant gauge with gauge parameter $\xi$, where the sum
runs over the transverse and longitudinal modes. 
The $P^{\mu\nu}_\alpha$ are the adequate projectors.
The corresponding 
self--energies are (see \cite{htl}):
\begin{equation}
\Pi_{_{T}}(L)=3m^2_{\hbox{\cmr g}}\left[{{l_0^2}\over{2l^2}}
+{{l_0(l^2-l_0^2)}\over{4l^3}}\ln\left({{l_0+l}\over{l_0-l}}\right)
\right]
\end{equation}
\begin{equation}
\Pi_{_{L}}(L)=3m^2_{\hbox{\cmr g}}\left[{{l^2-l_0^2}\over{l^2}}\right]
\left[1
-{{l_0}\over{2l}}\ln\left({{l_0+l}\over{l_0-l}}\right)
\right],
\end{equation}
where $m^2_{\hbox{\cmr g}}=g^2T^2[N+N_{f}/2]/9$.
 This resummed gluon
propagator possesses two mass shells above the light cone: a 
transverse and a longitudinal one. Therefore, everything happens as
if the gluon had acquired a kind of mass through thermal 
corrections, this
mass being of order $gT$. Concerning the effective 
quark propagator, 
we will need only its hard momentum limit:
\begin{equation}
S(P)={{i\slP}
\over{P^2-M^2_{f}+i\epsilon}},
\end{equation}
where $M^2_{f}=g^2C_{_{F}}T^2/4$ is a thermal mass 
of order $gT$ (we started
from massless bare quarks).

\setbox10=\vbox{\hbox to 15 cm{\hfill ENSLAPP-A-613/96}
\hbox to 15 cm{\hfill hep-ph/9608429}}
\ht10=0mm
\dp10=0mm
\wd10=0mm

To achieve gauge invariance in the effective theory
obtained by this resummation, we need also some effective vertices.
In actual calculations, they are simply replaced by their
diagrammatic expansion (another way to deal with
these vertices is to use the Ward identities verified by the
HTLs). For example, the effective $qqg$ vertex
 reads (where a solid dot denotes an effective vertex
or propagator):
\begin{center}
\leavevmode
\hbox to\textwidth{\epsfbox{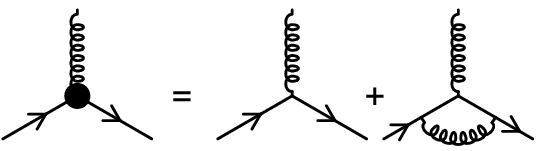}\hfill}
\end{center}
It is important to note that may exist some effective vertices without 
bare equivalent, like the $\gamma\gamma q q$ vertex:
\begin{center}
\leavevmode
\hbox to\textwidth{\epsfbox{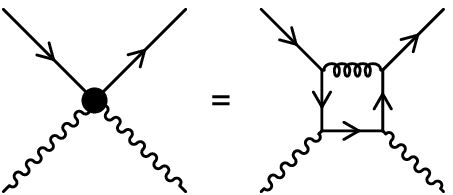}\hfill}
\end{center}

\subsection{Physical interpretation}
Physically, this resummation is closely related to the
Debye screening in a plasma. 
Indeed, if we look at the static limit of the self-energies 
$\Pi_{_{T,L}}$ for space-like gauge bosons, we may obtain
a mass: $m^2_{_{T,L}}\equiv \lim_{l\to 0} \Pi_{_{T,L}}(l_0=0,l)$.
The interaction range is
the inverse of this mass, and we have an exponential screening of
the field of a test charge put in the plasma. In QED, where exists
the classical model of Debye and Huckel, we can verify that the
mass $m_{_{L}}$ is precisely
 the classical Debye mass for electric fields,
whereas the nullity of $m_{_{T}}$ corresponds to the fact that static
magnetic fields are not screened.

\subsection{Subsidiary problems}
Even if this HTL resummation is carried on, there remain some 
residual divergences. 

The first kind of problems is related to the fact 
that static ``magnetic"
(transverse) fields are not screened in a plasma. This
fact is well known in a QED plasma, for which it has been
proven that the ``magnetic mass" $m_{_{T}}$ does vanish at all orders of
perturbation theory. For QCD and other non abelian gauge
theories, the status of this magnetic mass is not so clear;
nevertheless it is strongly expected that it is beyond the abilities
of perturbative methods and of order $g^2T$ or smaller. A long 
standing problem where this kind of transverse infrared 
divergence does appear 
is the calculation of the fermion damping rate. 

Other divergences which are not cured by the
HTL resummation, at least in its minimal version, are collinear
ones. Indeed, collinear divergences may occur whenever we have
massless particles in a diagram. Since in the standard 
HTL framework only soft lines are resummed and can therefore
acquire a thermal mass, we can still have problems with 
hard massless particles. If we come back to the circumstances
which necessitate a resummation, we see that we should compare 
the quantities $P^2$ and the self-energy $\Pi(P)={\cal O}(g^2T^2)$.
Of course, as said before, we need a resummation when the components
of $P$ are soft; but we need also effective propagators when
the components of $P$ are hard, but $P$ is close to the light--cone.
It has been shown recently in \cite{rebhan} that this supplementary 
resummation
still preserves gauge invariance. We will see in the next section
on the example of photon production that it can cure the collinear
divergences, and, unexpectedly, some transverse infrared divergences.

\section{SOFT PHOTON PRODUCTION}
\subsection{Various approaches}
\subsubsection{Semi-classical methods}
Recently, in order to compute the $\gamma$ emission rate by an
ultrarelativistic parton of the plasma, semi-classical methods
have witnessed a renewed interest. Basically, the idea of these 
approaches
is to look at the emission of a single photon by a quark undergoing
multiple scatterings while crossing the plasma. An essential
hypothesis of this method is to assume that the scattering centers
are static, so that only longitudinal bosons are exchanged in the
scattering process. Also, the interaction is assumed to be 
Debye screened. The essential physical result of this approach is
a suppression of the emission rate at small $\gamma$ energies,
known as the Landau-Pomeranchuk-Migdal (LPM) effect
\cite{lpm}. Indeed, the lifetime of a quark emitting a $\gamma$
of energy $\omega$ at an angle $\theta$
after a scattering is $\tau=1/\omega\theta^2$. If this 
lifetime is greater than the mean free path $\lambda$
of the quark in the medium,
then the $\gamma$ emission is suppressed because a new scattering
occurs before the $\gamma$ emission: this is the case when 
$\omega <1/\lambda$.

\subsubsection{Thermal field theory}
In thermal field theory, the emission rate per unit volume of the plasma
is related to the
imaginary part of the $\gamma$ self--energy by:
\begin{equation}
q_0{{dN}\over{d^3\vec{q}d^4x}}=-{1\over{(2\pi)^3}}n_{_{B}}(q_0) 
\hbox{\rm Im}\,\Pi^\mu{}_\mu(Q),
\end{equation}
the imaginary part being itself obtained by a generalization at finite 
$T$ of the standard cutting rules. In the standard HTL framework at
$1$ loop \cite{bpy}, we have three diagrams:
\begin{center}
\leavevmode
\hbox to\textwidth{\epsfbox{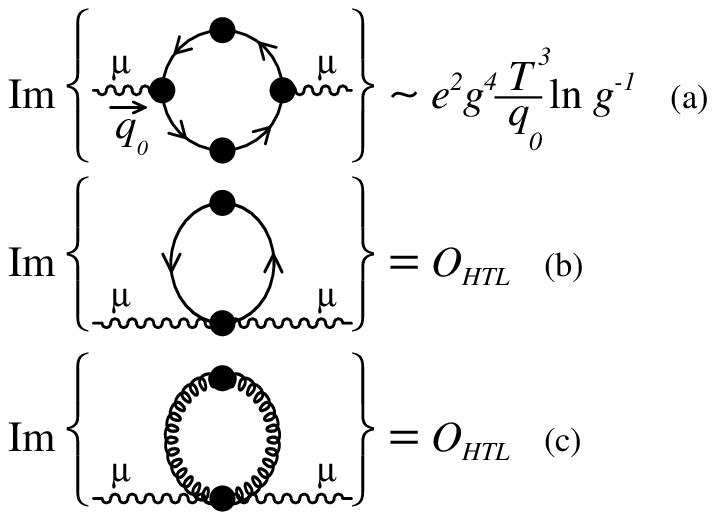}\hfill}
\end{center}
where the symbol $0_{_{HTL}}$ means that we obtain a null result
within the standard rules for the HTL expansion. Some of the relevant
corresponding processes are:
\begin{center}
\leavevmode
\hbox to\textwidth{\epsfbox{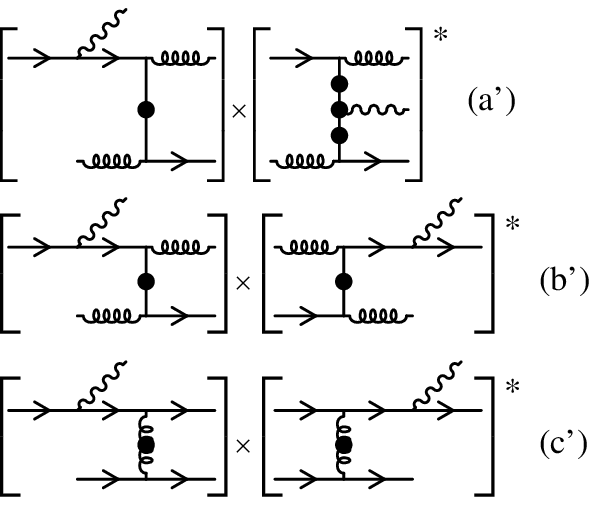}\hfill}
\end{center}
We can see that the dia\-gram $(a)$ in\-vol\-ves
the in\-ter\-fe\-ren\-ce of a ``bremss\-tra\-hlung-li\-ke" pro\-cess
with a very complicated one (see $(a')$), whereas the diagrams 
$(b)$ and $(c)$
involve only bremsstrahlung processes
(see $(b')$ and $(c')$). From a
physical point of view, it seems very surprising that bremsstrahlung
is subdominant in front of the complicated process of diagram $(a)$.
This was our motivation to study the last two diagrams beyond
the standard HTL scheme. Moreover, since the
parton exchanged in the scattering is soft and since in
that kinematical regime a bosonic statistical weight
dominates over a fermionic one, the diagram $(c)$ is
dominant with respect to $(b)$. 

\subsection{Bremsstrahlung diagrams}
The remaining of this
talk will be devoted to the study of the diagram $(c)$.
In order to do this calculation, we replace the $\gamma\gamma g g$
effective vertex by the corresponding loops, which gives:
\begin{center}
\leavevmode
\hbox to\textwidth{\epsfbox{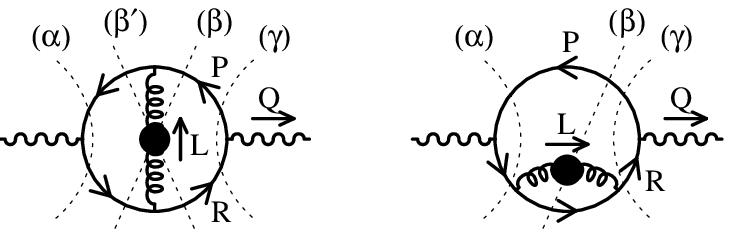}\hfill}
\end{center}
In fact, we have a third diagram, involving a self-energy
correction on the upper fermion line (not represented here).
Since we are interested in the production of photons with
a positive virtuality $Q^2\geq 0$ (which is the case of
interest for the production of real photons or of pairs
of leptons), the cuts $(\alpha)$ and $(\gamma)$ are
vanishing for kinematical reasons.
\subsubsection{Simplified form of ${\rm Im}\,\Pi^\mu{}_\mu$}
In order to calculate the contribution to ${\rm Im}\,\Pi^\mu{}_\mu$
of the previous two diagrams, we use the so-called 
Retarded/Advanced formalism \cite{formalism}. It appears that the
terms corresponding to the cut $(\beta)$ are very similar
in the two diagrams (the cut $(\beta')$, to be combined
with the non vanishing cut of the other self-energy correction, gives
a contribution equal to that of the cut $(\beta)$).
In our kinematical domain ($P,R\sim T$ and $Q,L\ll T$), we obtain:
$$\eqalign{
{\rm Im}\,\Pi^\mu{}_\mu(Q) \approx
 8 (-1)_{_{L}} e^2g^2 
\int{{d^4R\,d^4L}\over{(2\pi)^6}}
\times&\cr{{q_0}\over{T}} n_{_{F}}(r_0)(1-n_{_{F}}(r_0))
n_{_{B}}(l_0)
\rho_{_{T,L}}(l_0,l)\times&\cr}$$
$$\eqalign{
\displaystyle{{{r^2}\over{l^2}}\;
{{(L^2)^2\;\delta(P^2-M^2_{f})\;
\delta((R+L)^2-M^2_{f})}\over
{(R^2-M^2_{f})\;((P+L)^2-M^2_{f})}}},&\cno\cr
}$$
where $n_{_{B,F}}(\omega)=1/(\exp(\omega/T)\mp1)$ and
$\rho_{_{T,L}}$ is the imaginary part of 
the
longitudinal or transverse part of the effective gluon
propagator: $\rho_{_{T,L}}(l_0,l)=2 {\rm Im}\,
(L^2-\Pi_{_{T,L}}(L)+i\epsilon)^{-1}$. 
The symbol $(-1)_{_{L}}$ denotes an extra minus sign in the longitudinal
gluon exchange contribution.
It is straightforward
to verify that the gauge dependent part of this propagator
disappears when the self-energy and vertex corrections are combined.
 Two kinds
of contributions arise in $\rho_{_{T,L}}$, corresponding
to very different features of the analytic structure of
the effective propagator. The first contribution, arising at $L^2>0$,
is made of Dirac delta functions corresponding to
the thermal mass shells of the resummed propagator ({\it i.e.} 
from poles in the $l_0$ plane).
The second contribution to this
imaginary part comes from the logarithms in $\Pi_{_{T,L}}$ at $L^2<0$
({\it i.e.} from a cut in the $l_0$ plane) and corresponds
to the effect known as Landau damping (damping of a wave
 by absorption of virtual photons by quarks of the 
plasma). In fact, in our rate, the term containing the pole part
of the functions $\rho_{_{T,L}}$ corresponds to a production by
Compton effect, where a thermalized on-shell gluon
is absorbed by a quark which re-emits a photon, rather than 
bremsstrahlung. In $(a'),(b')$ and $(c')$ we represented 
only the amplitudes corresponding to 
bremsstrahlung.
Nevertheless, a careful analysis of the
Dirac delta constraints in $(6)$ 
shows that the region $L^2>0$ is forbidden
as far as $Q^2/q_0^2\ll1$, which is the situation we are interested in.
Therefore, the bremsstrahlung production dominates over the Compton one.

\subsubsection{Enhancement mechanism}
  Taking into account the constraints provided by the
two Dirac delta functions, we obtain for the denominators (in the
region where $P,R\sim T$ and
$Q,L\ll T$):
$$\eqalign{
\smash{R^2-M^2_f\approx 2q_0r\left[1-\cos\theta+{{M^2_{eff}}\over{2r^2}}
\right]}&\cno\cr
\int\limits_{0}^{2\pi} {{d\phi/2\pi}\over{(P+L)^2-M^2_f}}\approx
(2q_0r)^{-1}\times&\cr}$$
$$\eqalign{
\left[
\left(1-\cos\theta+{{M^2_{eff}}\over{2r^2}}+{{L^2}\over{2r^2}}\right)^2
\!\!\!-{{L^2 M^2_{eff}}\over{r^4}}\right]^{-1/2}&\cno\cr}$$
where $\theta$ is the angle between the spatial components of $R$ and
$Q$, $\phi$ is the azimuthal angle
between the spatial parts of $Q$ and $L$, 
and $M^2_{eff}\equiv M^2_{f}+Q^2r^2/q_0^2$ (we integrated over $\phi$ 
here since $(P+L)^2-M^2_f$ is the only place where $\phi$
appears). At that point, it is clear that the potential
collinear divergence at $\theta=0$ is regularized by a combination
of the fermion thermal mass and of the photon virtuality. At $Q^2=0$,
it appears to be essential to keep this fermion thermal mass,
despite the fact that the fermion has a hard momentum.

Na\"\i vely, we obtain the following order for the integral over 
$\cos\theta$:
\begin{equation}
I\equiv\int\limits_{-1}^{+1} {{d\cos\theta}\over
{(R^2-M^2_f)((P+L)^2-M^2_f)}}\sim {{1}\over{q_0^2 r^2}}
\end{equation}
In order to explain the enhancement mechanism,
we can roughly approximate:
$$\eqalign{
R^2-M^2_{f}\sim 2 q_0 r(1-\cos\theta+u^*)&\cr
(P+L)^2-M^2_{f}\sim 2 q_0 r(1-\cos\theta+\tilde{u}^*),&\cno\cr
}$$
where $u^*, \tilde{u}^* \sim M^2_{eff}/r^2$ and 
$u^*-\tilde{u}^*\sim L^2/r^2$,
so that we get:
$$\eqalign{
\hbox{If } u^*\leq L^2/r^2\qquad I\sim {{1}\over{u^*-\tilde{u}^*}}
{{1}\over{q_0^2r^2}}\gg {{1}\over{q_0^2r^2}}&\cr
\hbox{If } u^*\gg L^2/r^2\qquad I\sim {{1}\over{u^*}}
{{1}\over{q_0^2r^2}}.&\cno\cr
}$$
Therefore, we have an enhancement over the order one
expects na\"\i vely as far as $u^*\ll 1$, {\it i.e.} if
$Q^2/q_0^2\ll 1$ (this is the meaning of ``near the light-cone").
 The origin of this enhancement
lies in the residues of collinear singularities: 
potential collinear divergences, although regulated by a 
fermion thermal mass, are at the origin of the break-down
of the power counting rules of the HTL expansion, which fails
to handle properly collinear sensitive processes.
\subsubsection{Final expression of ${\rm Im}\,\Pi^\mu{}_\mu$}
Finally, after taking into account the $\delta-$cons\-traints and 
per\-for\-ming the angular integration, we obtain:
$$\eqalign{
{\rm Im}\,\Pi^\mu{}_\mu(Q)\approx (-1)_{_{T}} 
{{e^2 g^2}\over{\pi^4}}{{T^3}\over{q_0}}\times&\cr
\int\limits_{0}^{+\infty}\!\!dv\,v^2 {{1}\over{e^v+1}}\left[1-
{{1}\over{e^v+1}}\right]\times&\cr
\int\limits_{0}^{1}\!\!{{dx}\over{x}}\widetilde{I}_{_{T,L}}(x)
\int\limits_{0}^{+\infty}\!\!\!dw{{\sqrt{\displaystyle{{{w}
\over{w+4}}}}{\rm tanh}^{-1}
\sqrt{\displaystyle{{{w}\over{w+4}}}}}\over
{(w+\widetilde{R}_{_{T,L}}(x))^2+\widetilde{I}^2_{_{T,L}}(x)}}&\cno\cr
}$$
where we introduced some useful dimensionless quantities:
$$\eqalign{
\widetilde{R}_{_{T,L}}\!\!\!\equiv {\rm Re}\,\Pi_{_{T,L}}/M^2_{eff}\qquad
\widetilde{I}_{_{T,L}}\!\!\!\equiv {\rm Im}\,\Pi_{_{T,L}}/M^2_{eff}\hidewidth&\cr
v\equiv r/T\qquad w\equiv -L^2/M^2_{eff}\qquad x\equiv l_0/l.&\cno\cr
}$$
In the expression above, we have factorized out the parameters
fixing the dimension and the order of magnitude on the
first line, so that the remaining two factors are only a numerical
factor, which will be denoted by $J_{_{T,L}}$ in what follows.
\section{CONCLUSIONS}
Let us review some properties of the
bremss\-trah\-lung pro\-duc\-tion rate quoted before in $(12)$.

(i) The regions $R\ll T$ and $L\sim T$ are negligeable in the
integral, so that
our approximations based on these assumptions are safe.

(ii) The result is infrared finite, without the need of a
magnetic mass for the transverse gluon exchange. This feature
is very important for QED, where this magnetic mass is
known to be zero to all orders.
 At $Q^2=0$, if we look at the formal limit $M_f\ll 
m_{\hbox{\cmr g}}$, we obtain:
$$\eqalign{
J_{_{L}}\sim \ln(m_{\hbox{\cmr g}}/M_f)\qquad
J_{_{T}}\sim \ln^2(m_{\hbox{\cmr g}}/M_f).
&\cno\cr
}$$
One power of that logarithm, common to both the transverse and
longitudinal contributions, is due to the potential
collinear singularity. The additional power of that logarithm in the
transverse term is a remnant of an infrared divergence
(related to the absence of magnetic mass at this order), unexpectedly
screened by the fermion thermal mass.

(iii) At $Q^2=0$, the bremsstrahlung contribution is of order
${\rm Im}\,\Pi^\mu{}_\mu\sim e^2 g^2 T^3/q_0$ and therefore dominates
over the contribution of the soft fermion loop $(e^2g^4 T^3/q_0)$.

(iv) Let us end with a plot of $J_{_{T,L}}$ as a function of
$\log(Q^2/q_0^2)$, with $g=0.1$, $N=3$ and $N_{f}=3$:
\begin{center}
\leavevmode
\hbox to\textwidth{\epsfbox{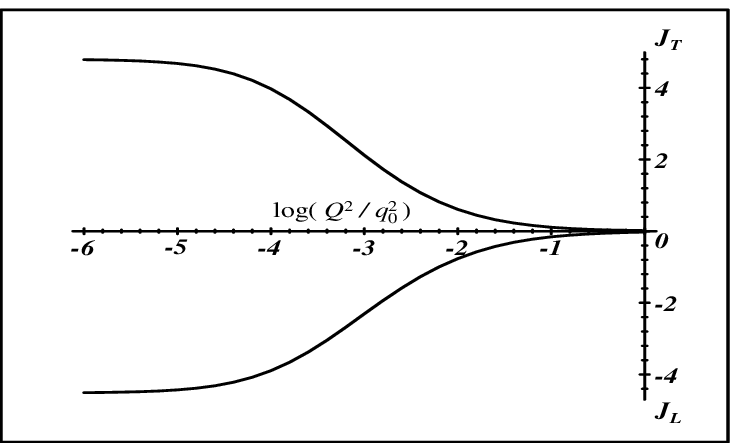}\hfill}
\end{center}
We note a flatness at small $Q^2$, reflecting the fact that
the mass $M_{eff}$ is dominated by the fermion thermal mass,
 and we see that the enhancement disappears if $Q^2/q_0^2\to 1$.

(v) The transverse and longitudinal contributions
are of the same magnitude, which seems to indicate that 
the static scattering centers hypothesis of the semi-classical
approach is wrong.

\vskip 3mm
\hbox to \linewidth{\bf Aknowledgements\hfill} 
It is a pleasure to thank P.~Aurenche for his critical reading of this
manuscript and S.~Narison for organising such a stimulating
conference.

\end{document}